\documentclass[conference]{IEEEtran}
\IEEEoverridecommandlockouts

\usepackage{scalerel}
\usepackage{cite}
\usepackage{amsmath,amssymb,amsfonts}
\usepackage{algorithmic}
\usepackage{graphicx}
\usepackage{textcomp}
\usepackage{xcolor}
\usepackage{subfig}
\usepackage[bottom]{footmisc}
\usepackage[bookmarks=false]{hyperref}

\begin{document}

\title{Elites Tweet? Characterizing the Twitter Verified User Network}

\author{
\IEEEauthorblockN{Indraneil Paul\IEEEauthorrefmark{1},
Abhinav Khattar\IEEEauthorrefmark{2},
Ponnurangam Kumaraguru\IEEEauthorrefmark{2}\IEEEauthorrefmark{4},
Manish Gupta\IEEEauthorrefmark{3} and
Shaan Chopra\IEEEauthorrefmark{2}
}

\IEEEauthorblockA{\IEEEauthorrefmark{1}International Institute of Information Technology, Hyderabad\\
\href{mailto:indraneil.paul@research.iiit.ac.in}{indraneil.paul@research.iiit.ac.in}}
\IEEEauthorblockA{
\IEEEauthorrefmark{2}Indraprastha Institute of Information Technology, Delhi\\
\href{mailto:abhinav15120@iiitd.ac.in}{abhinav15120@iiid.ac.in}, \href{mailto:pk.guru@iiit.ac.in}{pk@iiid.ac.in}, \href{mailto:shaan15090@iiitd.ac.in}{shaan15090@iiid.ac.in}
}
\IEEEauthorblockA{\IEEEauthorrefmark{3}Microsoft 
India\\
\href{mailto:gmanish@microsoft.com}{gmanish@microsoft.com}
}
}

\maketitle

\begin{abstract}
Social network and publishing platforms, such as Twitter, support the concept of \textit{verification}. Verified accounts are deemed worthy of platform-wide public interest and are separately authenticated by the platform itself. There have been repeated assertions by these platforms about verification not being tantamount to endorsement. However, a significant body of prior work suggests that possessing a verified status symbolizes enhanced credibility in the eyes of the platform audience. As a result, such a status is highly coveted among public figures and influencers. Hence, we attempt to characterize the network of verified users on Twitter and compare the results to similar analysis performed for the entire Twitter network. We extracted the entire network of verified users on Twitter (as of July 2018) and obtained 231,246 English user profiles and 79,213,811 connections. Subsequently, in the network analysis, we found that the sub-graph of verified users mirrors the full Twitter users graph in some aspects such as possessing a short diameter. However, our findings contrast with earlier findings on multiple aspects, such as the possession of a power law out-degree distribution, slight dissortativity, and a significantly higher reciprocity rate, as elucidated in the paper. Moreover, we attempt to gauge the presence of salient components within this sub-graph and detect the absence of homophily with respect to popularity, which again is in stark contrast to the full Twitter graph. Finally, we demonstrate stationarity in the time series of verified user activity levels. To the best of our knowledge, this work represents the first quantitative attempt at characterizing verified users on Twitter.
\end{abstract}


\section{Introduction}
All major social networking websites including Twitter, Facebook, and Instagram support the concept of \textit{verified accounts}\footnotemark[1]\footnotetext[1]{The exact term varies by platform, with Facebook using the term ``Verified Profiles''. However in the interest of consistency, all owner-authenticated accounts are referred to as \textit{verified accounts}, and  their owners as \textit{verified users}.\\ \IEEEauthorrefmark{4} This work was partially done by Prof. Ponnurangam Kumaraguru while in sabbatical at IIIT Hyderabad.}, wherein users are independently authenticated by the platform. This status is usually conferred to accounts of well-known public personalities and businesses and is indicated with a badge next to the screen name (e.g., \scalerel*{\includegraphics{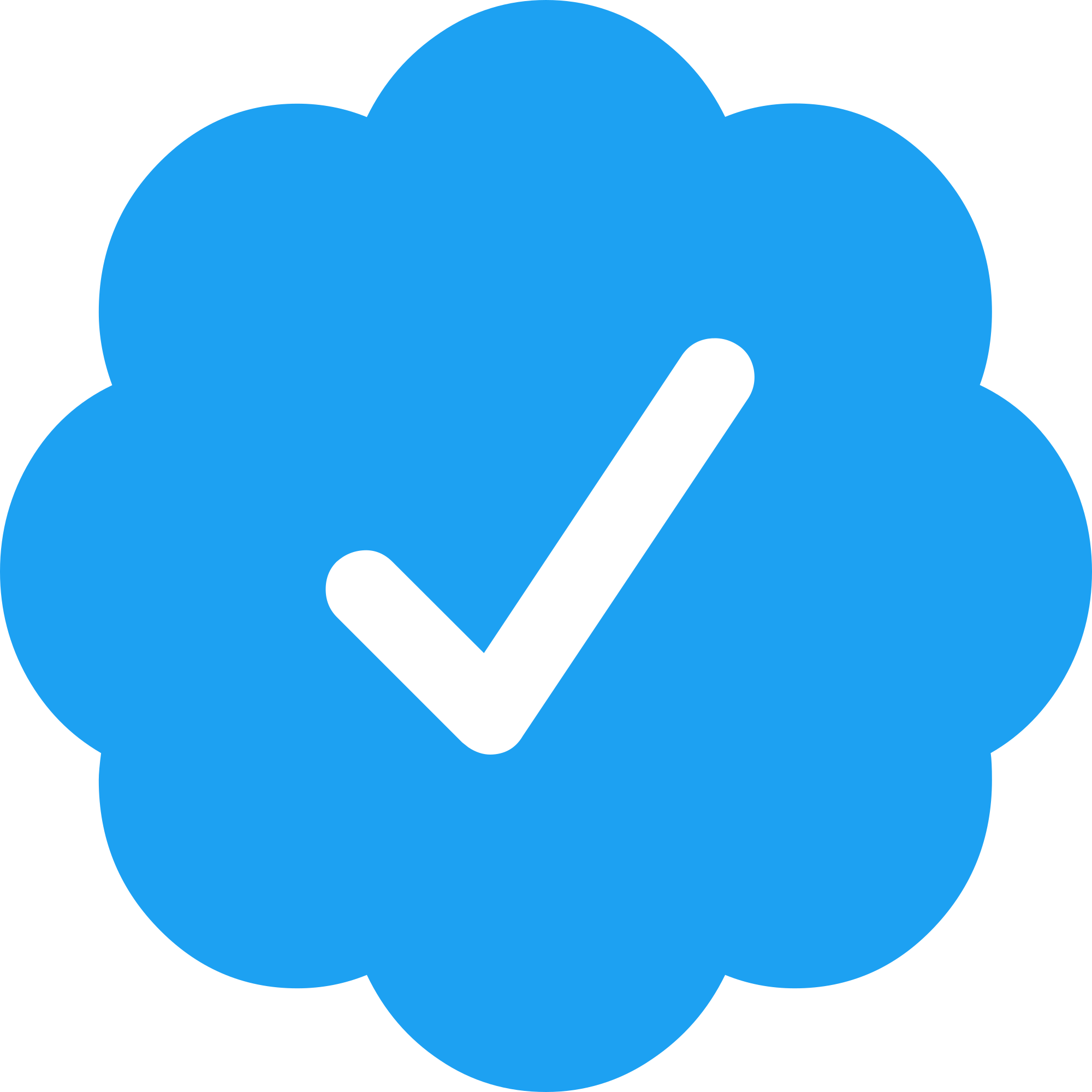}}{B} on Twitter and  \scalerel*{\includegraphics{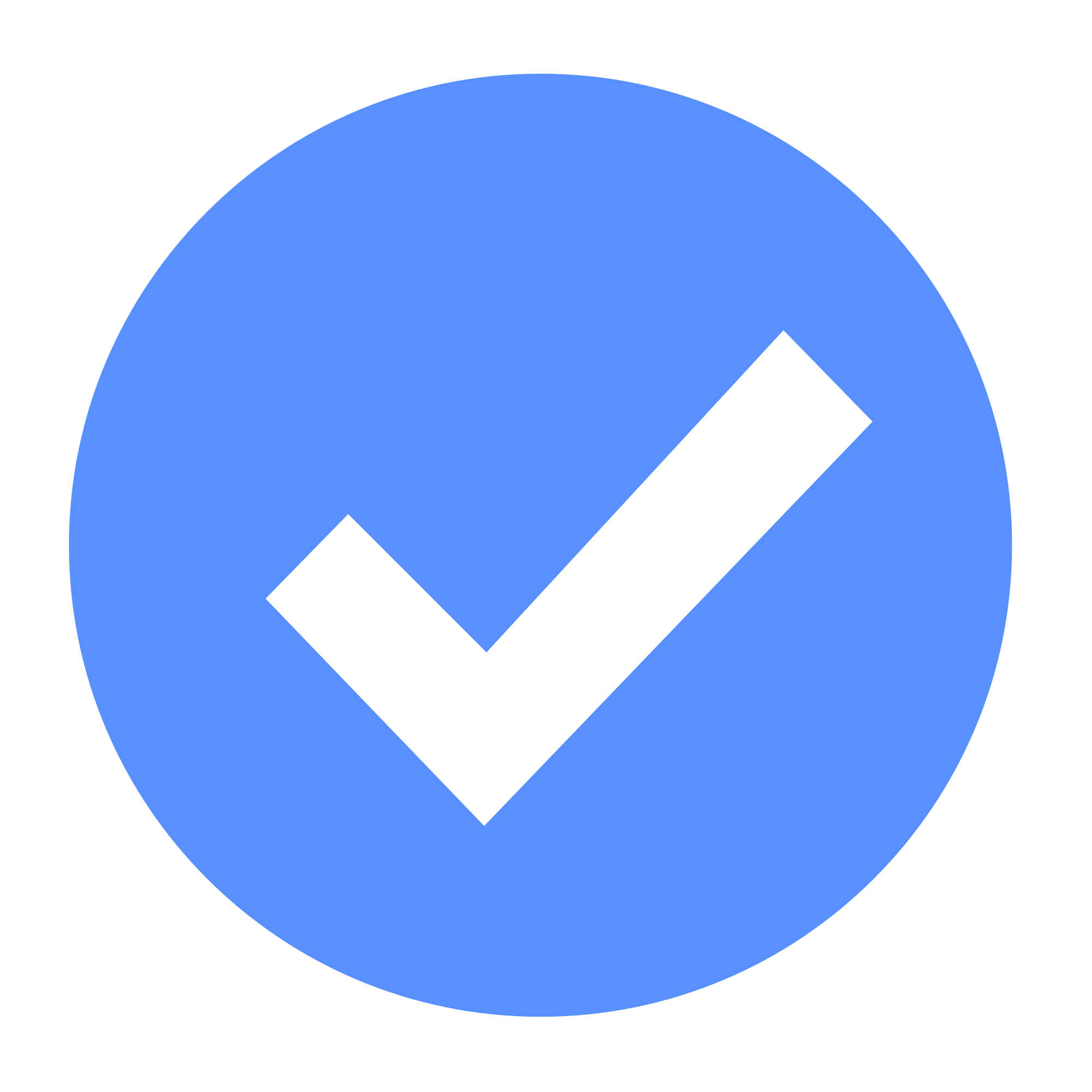}}{B} on Facebook).

For most of its existence, Twitter's own verification process has been an opaque one punctuated by sporadic attempts to open up the process to the general audience. Their verification policy~\cite{b9} states that an account is verified if it belongs to a personality or business deemed to be of sufficient public interest in diverse fields, such as journalism, politics, sports, etc. The exact intricacies of what they consider before verifying a handle are a trade secret. However, characterizing these users and gleaning into the ways in which the Twitter sub-graph induced by verified user nodes differ from the whole Twitter users network may yield insights into what discriminates verified users from non-verified ones.

\subsection{Motivation}
Despite repeated claims by Twitter that verification is not equivalent to accreditation, literature from social sciences~\cite{b2} and psychology suggests~\cite{b3} that the presence of a verified badge can add further credibility to the tweets made by a user handle. In addition, prior psychological tests~\cite{b15} have also revealed that the credibility of a message and its reception is influenced by its purported source and presentation rather than its pertinence or credulity. Existing work~\cite{b16} also indicates that widely endorsed information originating from a well-known source is easier to perceive as trustworthy. Owners of verified accounts are usually well-known and their content is on an average more frequently liked and retweeted than that of the generic Twittersphere~\cite{b37,b38}. 


Tweets pose a challenging scenario for credibility assessment owing to their limited length, negligible customization of visual design, and the frenetic pace at which they are consumed - with an average user devoting only three seconds of attention per tweet~\cite{b18}. Users may often resort to heuristics in order to judge online content. In~\cite{b4,b17}, heuristic-based models are presented for online credibility evaluation. Particularly relevant to this inquiry is the \textit{endorsement heuristic}, which is associated with credibility conferred to it (e.g. a verified badge) and the \textit{consistency heuristic}, which stems from endorsements by several authorities (e.g. a user verified in one platform is likely to be verified on others).

In the presence of aforementioned evidence along with pervasive fake content in the Twittersphere, our work explores how possessing a verified status can make a difference in outreach/influence of a brand or individual in terms of the extent and quality. Characterizing these ``elite'' users in isolation represents the first step in understanding how to become one of them. 

The rest of the paper is organized as follows. Section II details relevant prior work, hence putting our work in perspective. Section III elaborates our data acquisition methodology. In Section IV and V, we conduct network and activity analysis on verified users, respectively. We conclude with Section VI. 

\section{Related Work}
In this section, we delve into previous work on social network analysis and verified accounts.

Kwak et al.~\cite{b6} were among the first to study Twitter with the aim of understanding its role on the web - whether it was better approximated as a traditional social networking site or a news source. They show how Twitter is a powerful network that can be used to study online human behavior and also report the ways in which an online social network such as Twitter differs from human social networks. Castillo et al.~\cite{b1} attempt to identify credible tweets based on a variety of profile features including whether the user was authenticated by the platform or not. Morris et al.~\cite{b2} examined factors that influence profile credibility perceptions on Twitter. They found that possessing an authenticated status is one of the most robust predictors of positive credibility. Semertzidis et al.~\cite{b30} looked into the content of user biographies on Twitter and studied the presence of homophily with respect to their topical content.

Chu et al.~\cite{b11} used a slew of features to identify Twitter handles that are generating automated content. One of the crucial features for their analysis was the presence of a verification badge. Along similar lines, Hentschel et al.~\cite{b8} assert that most non-verified users on Twitter are within 7 degrees of separation of a verified user and a large majority of spam handles are located within 7-10 degrees of separation from verified users. This finding is promising with respect to fighting spam on Twitter as it suggests a white-listing mechanism for maintaining a core of non-spam users who are within a few degrees of separation of verified users.  

Java et al.~\cite{b13} analyzed Twitter's user base and activity-level growth. They studied the geographical distribution of users along with graph-based inquiries such as network reciprocity and region-localized clustering coefficients. Wang et al.~\cite{b12} uncovered influential handles on Twitter and proposed a metric as an improvement to the Topical PageRank employed by Twitter at the time. A positive correlation of PageRank with conventional metrics of the extent of influence, such as the number of followers, has been found for the entire Twittersphere~\cite{b6}. However, such an attempt has not yet been made for the network internal to the verified users on the platform. A large number of reciprocal connections on the Twitter network can be explained by \textit{homophily} of topical interests. Whether verified users, too, form reciprocal network links based on the same underlying principle was yet to be explored. 

Hence, to summarize, there exists a rich body of literature studying the characterization of users in the entire Twitter network. However, none of them, to the best of our knowledge, have attempted to characterize the Twitter sub-graph induced by the verified users. To that end, we run a battery of tests on the extracted network of verified users in order to uncover how this sub-graph behaves in comparison to the entire Twitter network.

\section{Dataset}

The `@verified' handle on Twitter follows all accounts on the platform that are currently verified. We queried this handle on the 18\textsuperscript{th} of July 2018 and extracted the IDs of 297,776 users who were verified at the time. Twitter provides a REST Application Programming Interface (API) with various endpoints that make data retrieval from the site in an organized manner easier. We used the REST API to acquire profile information of the verified user handles obtained previously. We further extracted a subset of verified users who had English listed as their profile language. This left us with 231,246 English verified users. Additionally, we leveraged a commercial Twitter Firehose in order to acquire a rich set of peripheral features of verified users.


For each verified user, we also queried the API in order to obtain the list of outlinks or friends of users that belonged to the aforementioned English subset. We filtered this list of friends and retained only those nodes that were leading to other verified users, thus obtaining the internal network existing among the verified users. The final network was an extremely sparse one with a density of 0.00148 and 231,246 English verified users having 79,213,811 directed-edges between them. The network consists of only 6027 isolated users with an average out-degree of 342.55 and a maximum out-degree of 114,815. The network had a notable giant strongly-connected component of 224,872 users which accounts for 97.24\% of the total English verified users. In all, the network contains 6251 connected components.  

Leveraging our access to the commercial Twitter Firehose, we obtained fine-grained time series of various user statistics, such as the number of followers, friends, and tweets, in the one year period of June 2017 to May 2018. This allowed us to calculate accurate averages and gauge growth trajectories of user reach and activity levels over time. We intend to anonymize and make this dataset public, once we have pursued all our inquiries to their logical end.

\begin{figure*}
\centering
\subfloat[][\textbf{Log Scaled Number of Users vs Friends}]{\includegraphics[width=0.49\textwidth,keepaspectratio]{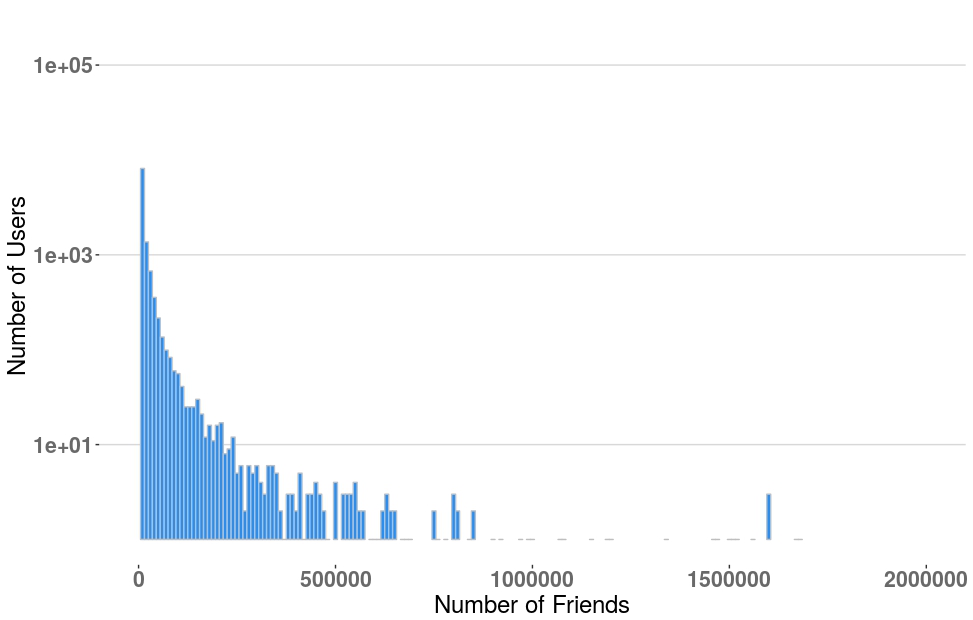}}\hspace{.005\linewidth} 
\centering
\subfloat[][\textbf{Log Scaled Number of Users vs Followers}]{\includegraphics[width=0.49\textwidth,keepaspectratio]{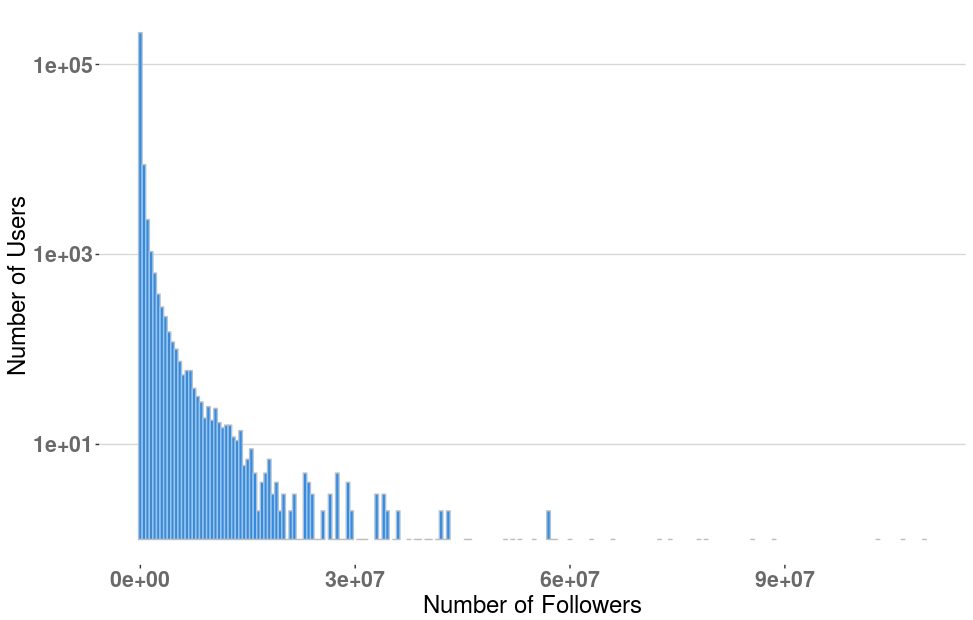}}\\
\centering
\subfloat[][\textbf{Log Scaled Number of Users vs List Memberships}]{\includegraphics[width=0.49\textwidth,keepaspectratio]{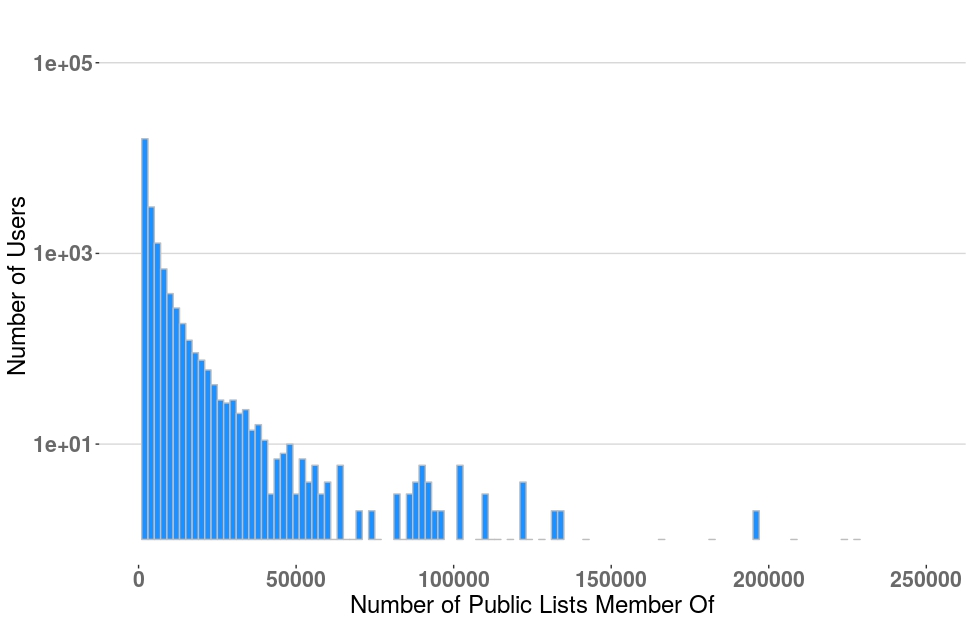}}\hspace{.005\linewidth} 
\centering
\subfloat[][\textbf{Log Scaled Number of Users vs Status Count}]{\includegraphics[width=0.49\textwidth,keepaspectratio]{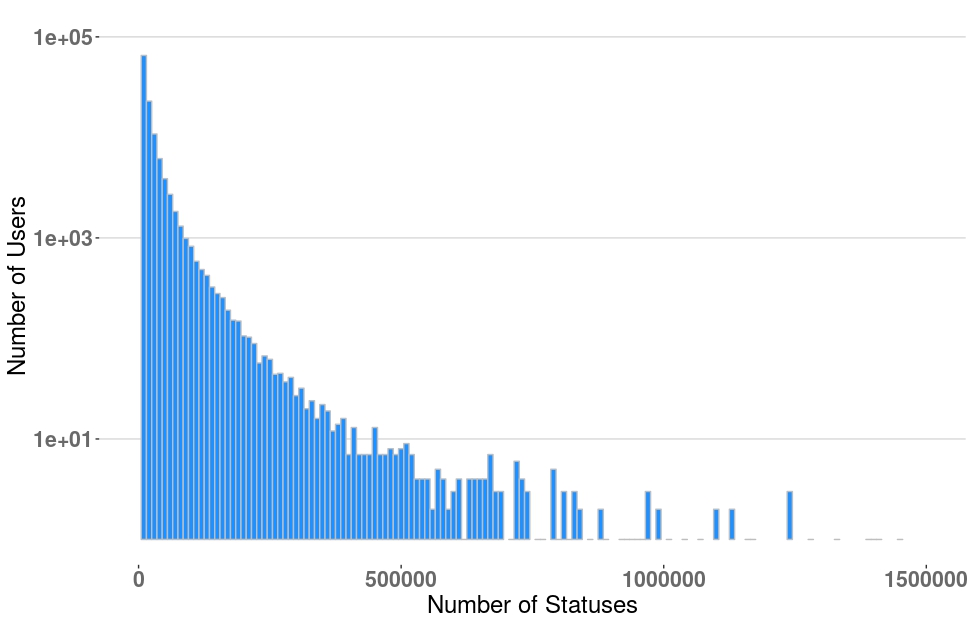}}\\
\centering
\caption{\textbf{Distribution of Friends, Followers, Public List Memberships and Tweet Activity.}}
\label{fig:fig1}
\end{figure*}

\section{Network Analysis}

We attempt to quantify how our network of verified users with English as their primary language, when considered in isolation, differs from the entire network. We analyze and compare our results to previous work on the Twitter network in its entirety.


\subsection{Basic Analysis}

The extracted network graph exhibits a very low density but a high level of connectedness. Out of the 231,246 English verified user nodes only 6027 are isolated, making the minimum out-degree 0, while a majority of the users belong to a single giant connected component. The greatest out-degree is 114,815 for the handle of a social media influencer - `@6BillionPeople'. The average out-degree is 342.55. The low density of the verified user network, as mentioned in the previous section, is further confirmed by a low average local clustering coefficient of 0.1583. The network has a slight degree dissortativity of -0.04 which is in contrast to the degree homophily formerly observed for the entire Twitter network~\cite{b6} and social networks in general~\cite{b7}. This suggests the existence of a large number of one-way relationships between prominent and semi-famous (medium degree) personalities which is further reinforced by the presence of 6091 attracting components (components in which if a random walk enters, it never leaves) in the directed graph. At the core of these components lie famous personalities (high in-degree users) who do not follow any other handle. These include handles of popular culture outlets such as `@ladbible', Hollywood screenwriters such as `@MrRPMurphy', and world-renowned spiritual gurus such as `@SriSri'. Figure~\ref{fig:fig1} displays the distributions of certain user metrics within the sub-graph of verified users.

\subsection{Degree and Eigenvalue Distribution}

Power-law is a key component in characterizing degree distribution of networks gathered from the world wide web and other large information sources. It is one of the hallmarks in the study of web-graphs and social networks~\cite{b20}. This matured into a series of theoretical studies about the presence of power-laws in other aspects of network structure such as eigenvalues of the Laplacian~\cite{b21,b22}. However, Kwak et al.~\cite{b6} reported an absence of a power-law in degree distribution when analyzing the whole Twitter network. This stands in contrast to our findings of a power-law almost entirely accounting for the out-degree distribution in the network of verified users. It also falls in line with existing work~\cite{b26} that identifies the presence of emergent properties observed in sampled sub-graphs and not seen in the graph as a whole. Subsequent work~\cite{b19} though has largely confirmed the presence of power-law in degree and Laplacian eigenvalue distributions of several synthetic and real world undirected social network datasets.

\begin{figure}[!bht]
\centerline{\includegraphics[width=0.95\linewidth,height=5.5cm]{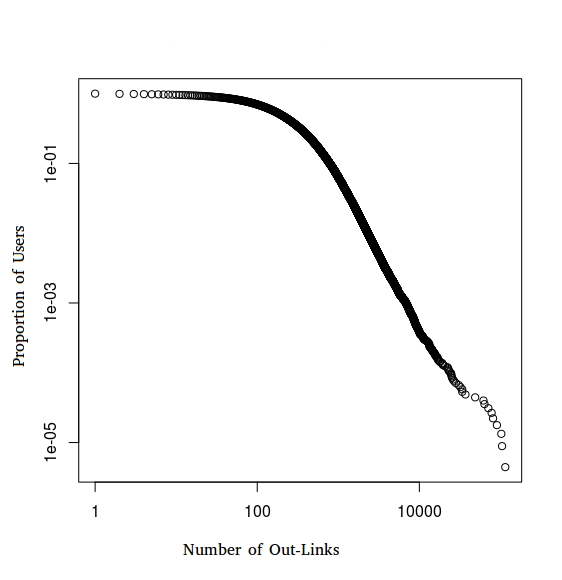}}
\caption{\textbf{Log-Log Scaled Distribution of Proportion of Users to Out-Degree}}
\label{fig:fig2}
\end{figure}

We computed the out-degree distribution as well as the largest 10,000 eigenvalues of the Laplacian matrix of the sub-graph. We discarded most of the smaller eigenvalues as the sparsity of our sub-graph resulted in most of those eigenvalues being close to zero, which could have caused floating point operation issues while inferring power-law. The eigenvalues were computed using the power iteration method in existing solvers. For both these distributions, we seek to compute the exponent $\alpha$ and a $x\textsubscript{min}$ threshold, which represents the lower bound of the best-fit range.  

Inferring of power-law parameters $\alpha$ and $x\textsubscript{min}$ is done using the maximum-likelihood algorithm by Clauset et al.~\cite{b25}. This approach is considered more accurate as compared to the traditional method of fitting the slope of the log-log plot. For the degree distribution, we use discrete maximum likelihood estimate (MLE) while for the eigenvalue distribution we use continuous MLE. We employ, the particular implementation by Nepusz~\cite{b23} which uses the BFGS algorithm to estimate the parameters. Moreover, this method and software calculate a goodness-of-fit parameter $p$ that indicates whether the power-law fit is likely to be significant. This score is based on a randomized procedure. If the value $p > 0.1$, then there is strong evidence that the presence of a power-law is justified. 

Continuous MLE inference for the eigenvalues yields parameter estimates of 3.18 for $\alpha$ and 9377.26 for $x\textsubscript{min}$ with a $p$ value of 0.3, thus indicating a very strong fit. Discrete MLE inference for the degree distribution yields parameter estimates of 3.24 for $\alpha$ and 1334 for $x\textsubscript{min}$ with a $p$ value of 0.13 indicating a significant fit. However, the closeness of the degree $p$ value to the threshold of 0.1 prompts us to conduct further pairwise tests to rule out other heavy-tailed distributions. We use an R toolbox~\cite{b24} to perform a Vuong's likelihood-ratio test between a power-law fit and alternate candidates such as log-normal, poisson and exponential fits. In each case, the tests returned significantly high 2-3 digit likelihood-ratio values indicating that the power-law was, in fact, the heavy-tailed distribution that best approximated the out-degree distribution in our sub-graph. The relationship between out-degree values and the proportion of users possessing it can be seen in Figure~\ref{fig:fig2}.

\subsection{Reciprocity}

The reciprocity rate refers to the proportion of pairs of links that go both ways. Kwak et al.~\cite{b6} have previously reported a reciprocity rate of 22.1\% among the directed links in the entire Twitter network. The verified user network has a significantly higher reciprocity rate of 33.7\%. This is still much lower than what is observed in other well-known social networks such as Flickr (68\%)~\cite{b35}. The likely cause of this is that entities like brands and third-party sources of curated and crawled information, which typically do not reciprocate engagements, are likely to be over-represented on Twitter. We conjecture, that the larger reciprocity rate viz-a-viz the whole Twitter graph is due to a larger core of publicly relevant and consequential personalities within this sub-graph. We leave validating this assertion for future work. 

\subsection{Degrees of Separation}

Ever since Stanley Milgram's seminal work on the ``Six Degrees of Separation''~\cite{b36}, the concept of using the distribution of pairwise node distances to characterize a social network has become commonplace. Watts et al.~\cite{b29} coined their small-world model after finding that many social and technological networks possessed small average path lengths. Prior work~\cite{b27} on an MSN messenger network of 180 million users revealed a median separation of 6 and an effective diameter (90 percentile path length) of 7.8.

\begin{figure}[!h]
\centerline{\includegraphics[width=0.95\linewidth,height=5.5cm]{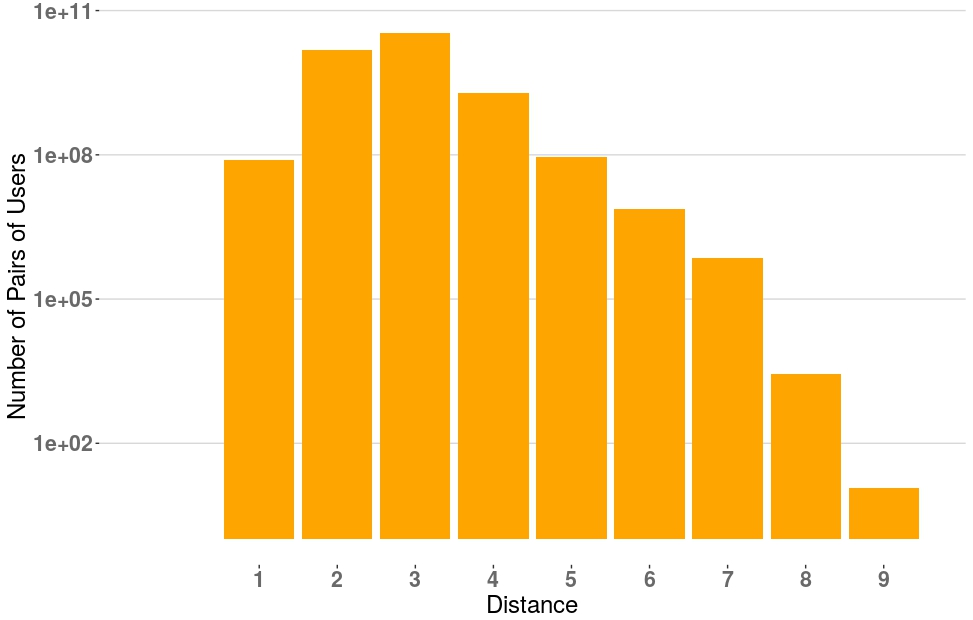}}
\caption{\textbf{Log Scaled Distribution of Number of Node Pairs vs Degrees of Separation.}}
\label{fig:fig3}
\end{figure}

The network of verified users on Twitter differs from the aforementioned networks as it is a network with directed edges. Thus, by notions of conventional graph-theoretic wisdom, one would expect the average path lengths to be higher as a path taken from a node to another need not be viable the other way. However, our analysis reveals that the average node distance to be 2.74, after omitting isolated nodes. Such a low number is especially surprising, given that the reciprocity rate is much lower compared to even other directed networks like Flickr. This value is considerably lower than the value of 4.12 reported for the general Twittersphere through a sampling mechanism~\cite{b6}. The distribution of pairwise node distances in the English verified sub-graph can be seen in Figure~\ref{fig:fig3}.

Later work~\cite{b28} using a bounded bi-directional search approach, optimally found the value of the average shortest path length on Twitter to be 3.43. This is still considerably higher than that of the verified sub-graph and reinforces the finding that while the Twitter verified sub-graph is sparse in its own right, it is still significantly denser than the whole of the Twitter graph at large.

\subsection{Verified User Bios}

Each user on Twitter can have a biography (or bio) allowing him/her to describe themselves using a limited number of characters. We attempt to gain insights from some of the most popular unigrams, bigrams and trigrams occurring in the bios of verified users. We also filter out n-grams constituted largely of non-informative words.  

\begin{figure}[!h]
\centerline{\includegraphics[width=0.95\linewidth,height=6cm]{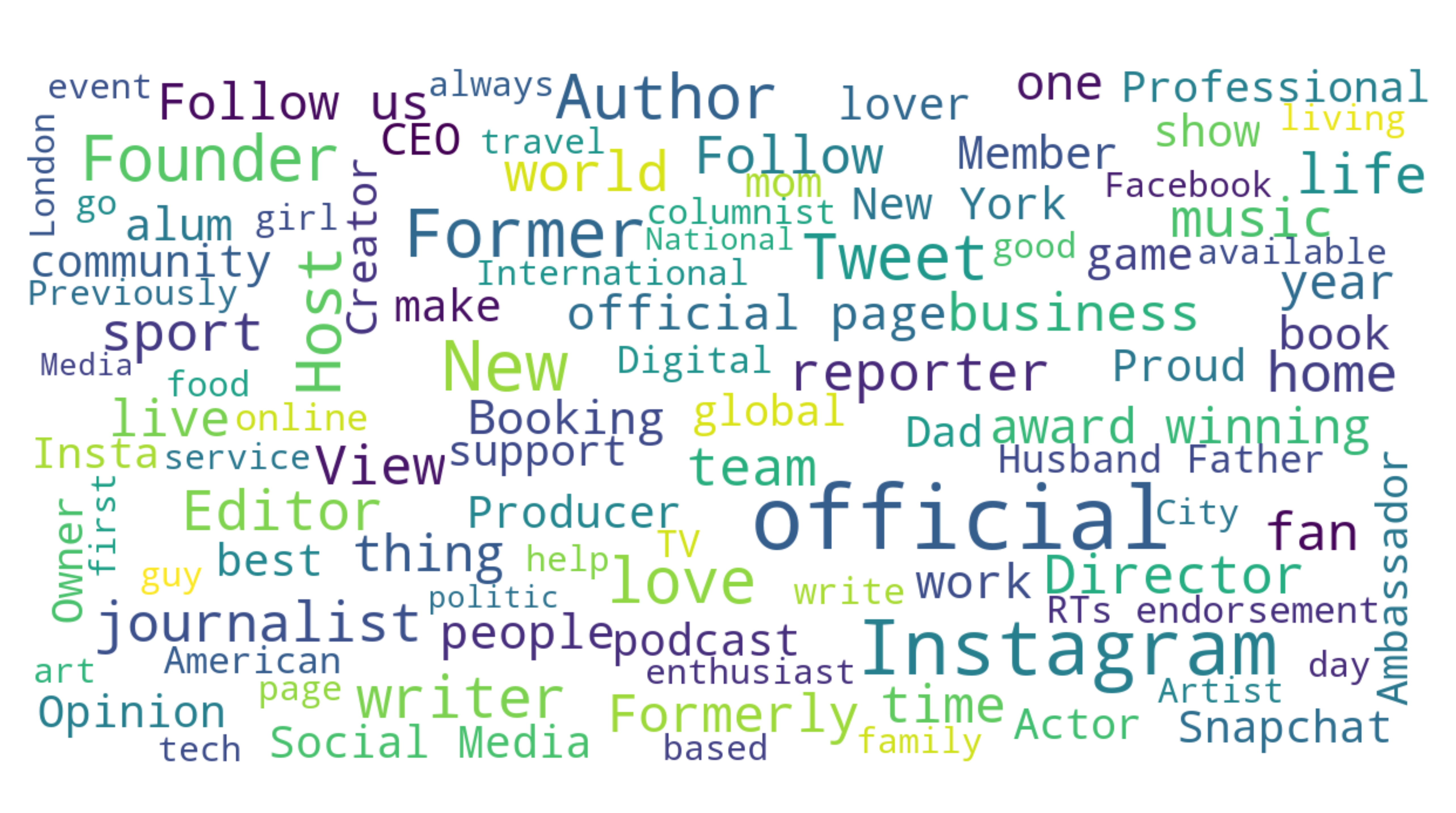}}
\caption{\textbf{Wordcloud of Most Frequent Unigrams in Bios of Verified Users.}}
\label{fig:fig4}
\end{figure}

The most frequent unigrams portray several underlying themes. They include cross-links to other social media accounts of an entity (\textit{``Instagram''}, \textit{``Facebook''} and \textit{``Snapchat''}), personal descriptors (\textit{``Husband''}, \textit{``Father''} and \textit{``Gay''}), professional descriptors (\textit{``Producer''}, \textit{``Founder''}, \textit{``Director''}, \textit{``Tech''}, \textit{``Author''} and \textit{``Sport''}), and terms relevant to businesses and brands online (\textit{``Booking''}, \textit{``Support''}, \textit{``International''} and \textit{``Official''}). Some unigrams such as \textit{``American''} and \textit{``London''} also hint towards the most dominant source of activity in the Anglospheric Twitter. Figure~\ref{fig:fig4} illustrates a word cloud of most frequent unigrams.

Bigrams and trigrams reiterate a largely similar narrative, dominated by generic descriptors (\textit{``Official Account''} and \textit{``Official Twitter Page''}), accomplishment descriptors (\textit{``Award Winning''}, \textit{``Olympic Gold Medalist''} and \textit{``Best Selling Author''}), professional descriptors (\textit{``Singer Songwriter''} and \textit{``Professional Rugby Player''}), and business and community-related terms (\textit{``Report Crimes Here''}, \textit{``Monday to Friday''} and \textit{``Weather Alerts EN''}). The most frequent bigrams and trigrams along with their respective frequencies can be seen in Table~\ref{tab:tab1} and Table ~\ref{tab:tab2}.

A running theme common to all three cases is the dominance of journalists and news and weather outlets. Several most frequent unigrams (\textit{``Journalist''}, \textit{``Reporter''} and \textit{``Editor''}), bigrams (\textit{``Breaking News''} and \textit{``Anchor Reporter''}), and trigrams (\textit{``New York Times''}, \textit{``Wall Street Journal''} and \textit{``Editor in Chief''}) are apropos of journalism. Being a pre-eminent journalist in an English media outlet seems to be one of the surest ways to get verified on Twitter. 

\begin{table}[htbp]
\caption{Most Popular Bigrams in Bios of Verified Users}
\begin{center}
\begin{tabular}{c|c}
\textbf{Bigram} & \textbf{Occurrences} \\
\hline
Official Twitter&12166  \\
Official Account&2788 \\
Award Winning&2270 \\
Follow Us&2268 \\
Co Founder&1581 \\
Husband Father&1540 \\
Opinions Own&1222 \\
New Album&1088 \\
Singer Songwriter&1043 \\
Co Host&933 \\
Latest News&904 \\
Breaking News&898 \\
Anchor Reporter&855 \\
Rugby Player&799\\
Managing Editor&769 \\
\hline
\end{tabular}
\label{tab:tab1}
\end{center}
\end{table}

\begin{table}[htbp]
\caption{Most Popular Trigrams in Bios of Verified Users}
\begin{center}
\begin{tabular}{c|c}
\textbf{Trigram} & \textbf{Occurrences} \\
\hline
Official Twitter Account&5457  \\
Official Twitter Page&1774 \\
Weather Alerts EN&847 \\
Emmy Award Winning&475 \\
New York Times&464 \\
Editor in Chief&461 \\
Best Selling Author&296 \\
Professional Rugby Player&253 \\
Wall Street Journal&252 \\
Professional Baseball Player&241 \\
Report Crime Here&238 \\
Award Winning Journalist&223 \\
For Customer Service&174 \\
Olympic Gold Medalist&174\\
Monday to Friday&174 \\
\hline
\end{tabular}
\label{tab:tab2}
\end{center}
\end{table}

\subsection{Centrality}

\begin{figure*}
\centering
\subfloat[][\textbf{List Memberships vs Betweenness Centrality}]{\includegraphics[width=0.49\linewidth,height=4.75cm]{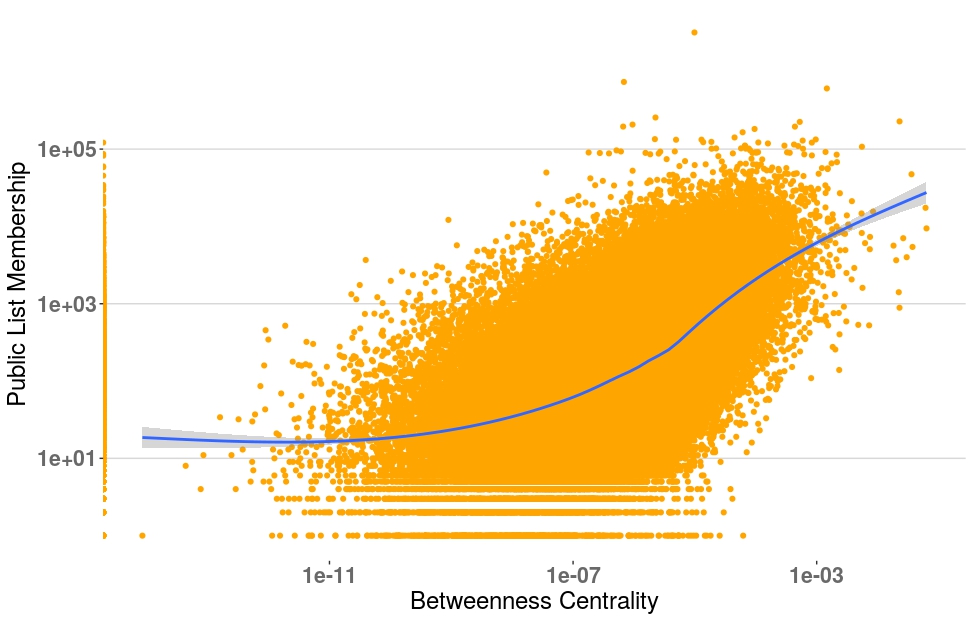}}\hspace{.005\linewidth} 
\centering
\subfloat[][\textbf{Follower Count vs Betweenness Centrality}]{\includegraphics[width=0.49\linewidth,height=4.75cm]{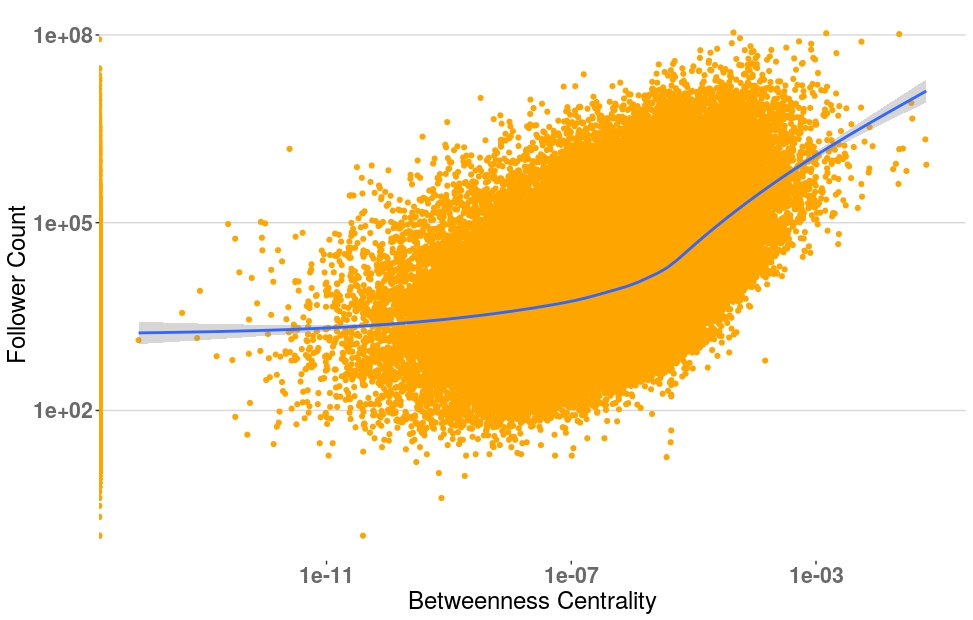}}\\
\centering
\subfloat[][\textbf{List Memberships vs PageRank Centrality}]{\includegraphics[width=0.49\linewidth,height=4.75cm]{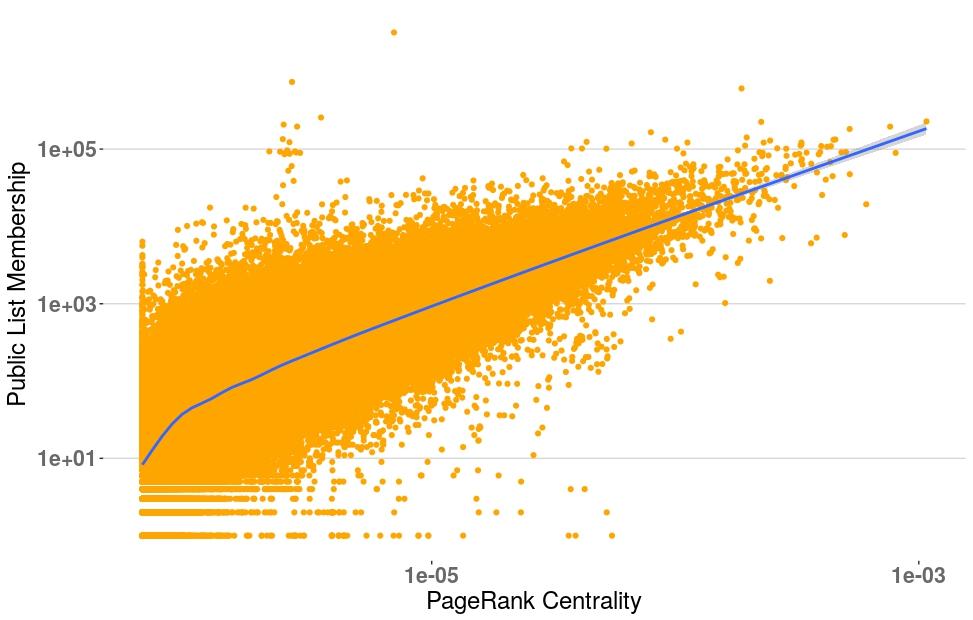}}\hspace{.005\linewidth} 
\centering
\subfloat[][\textbf{Follower Count vs PageRank Centrality}]{\includegraphics[width=0.49\linewidth,height=4.75cm]{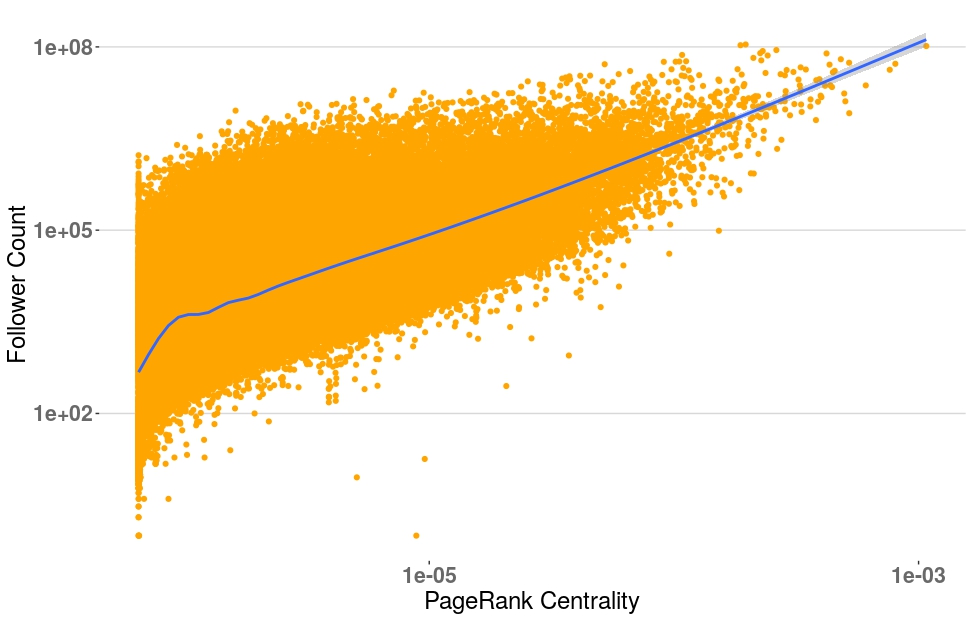}}\\
\centering
\subfloat[][\textbf{Follower Count vs Status Count}]{\includegraphics[width=0.49\linewidth,height=4.75cm]{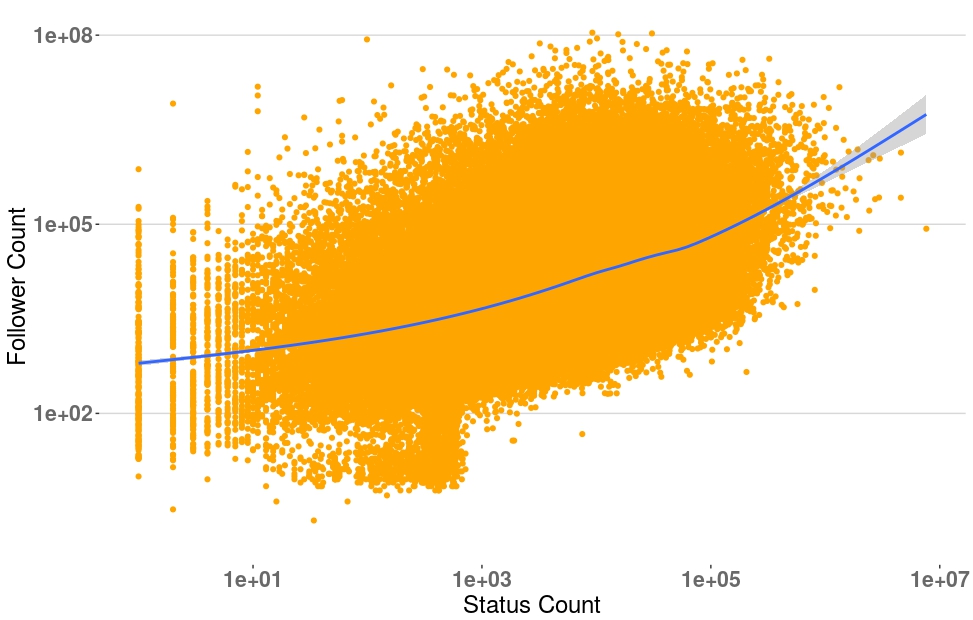}}\hspace{.005\linewidth} 
\centering
\subfloat[][\textbf{Follower Count vs List Memberships}]{\includegraphics[width=0.49\linewidth,height=4.75cm]{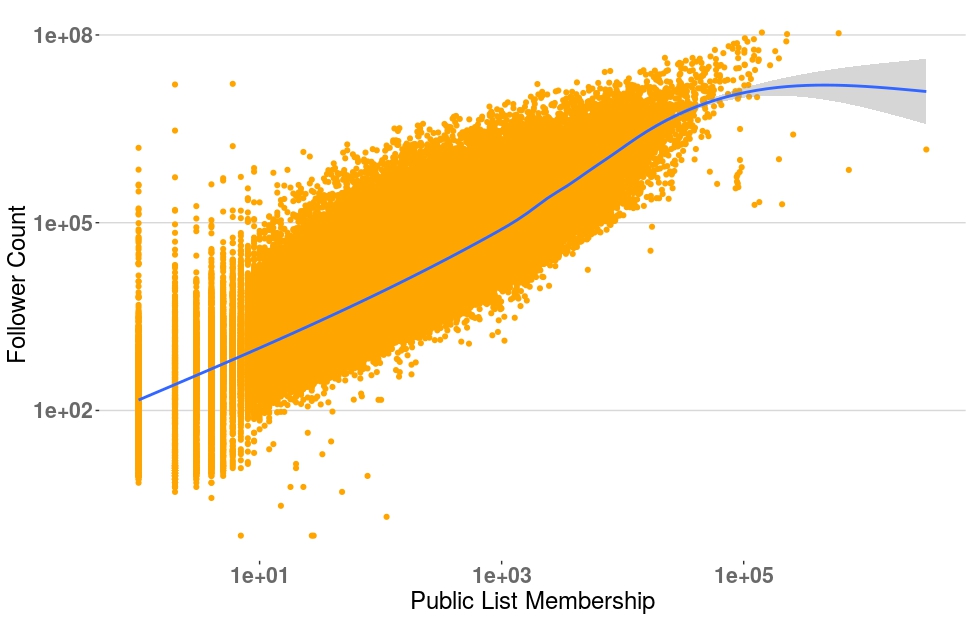}}\\
\centering
\caption{\textbf{Log-Log Scaled Scatter Plots of Various Influence Measures. The Regression Splines and 95\% Confidence Intervals are computed Using a Generalized Additive Model.}}
\label{fig:fig5}
\end{figure*}

To gain a better understanding of verified users we investigate how various centrality measures correlate with one another. These observations are illustrated in Figure~\ref{fig:fig5}. We study the relationship between the number of tweets made by a user and his/her followers. We observed that, in this aspect, the English verified sub-graph behaves exactly like the entire Twitter graph as previously reported in~\cite{b6}; the number of followers is seen trending upwards with an increase in the number of statuses and this trend becoming more apparent at higher extremes. Next, we look into the variation of the number of followers of a user with respect to the number of public lists the user is a part of. List membership has been shown to be a robust predictor of influence and topical relevance on Twitter~\cite{b31}. It has shown a competitive performance in topically recommending influential users to follow. This is further reinforced by our observation that the number of followers a user has, almost exclusively trends upwards with an increase in the number of list memberships. Diminishing returns set in only in the very upper echelons. 

\begin{figure*}[!htb]\centering
\includegraphics[width=\linewidth,height=8cm]{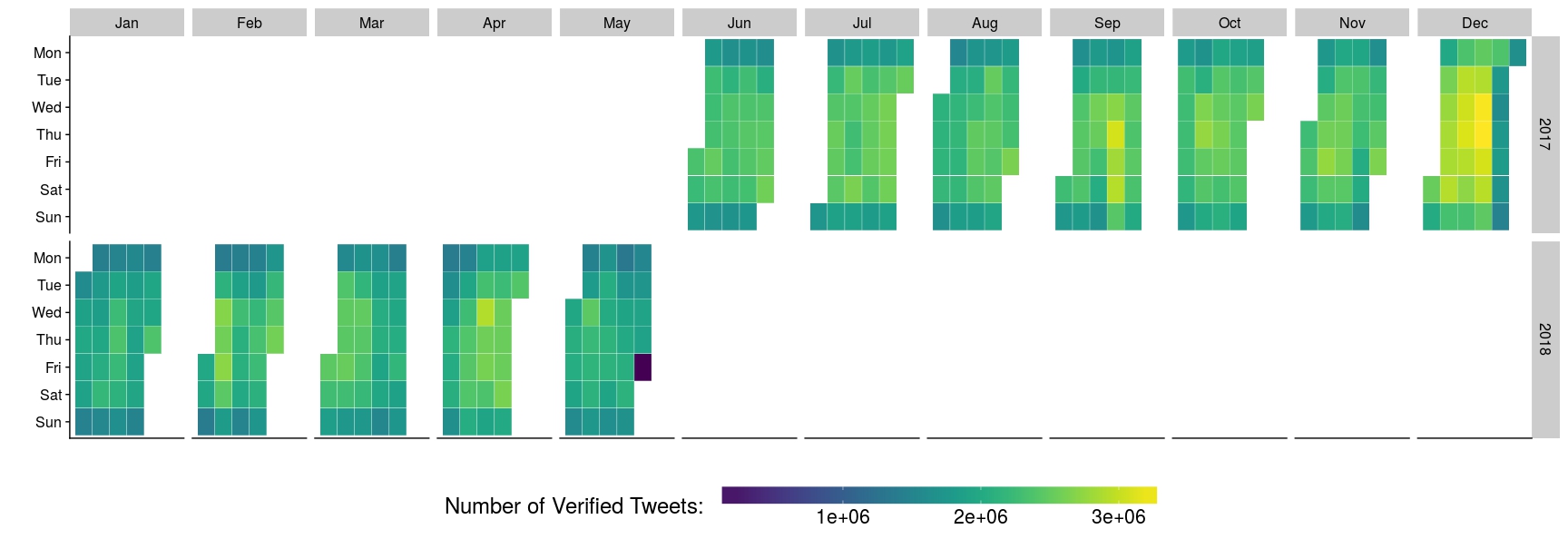}

\caption{\textbf{Calendar Maps for Verified User Tweet Activity Levels Over our One-Year Collection Period}}
\label{fig:fig6}
\end{figure*}

As claimed by Twitter's guiding principle, a user is verified only when he/she is deemed to be of sufficient public interest. We posit that Betweenness and PageRank centrality of a user in the sub-graph of English verified users can predict his/her reach in the overall network, such as the followers count. On testing, we observed that public list membership and follower count in the entire Twitter network is positively correlated with PageRank and Betweenness of that user in the English verified user sub-graph. In particular, the relationship between PageRank and total followers and list memberships were especially strong. Even though the correlation between follower count and Betweenness seems lukewarm at first, a strong relationship emerges at the higher extremes. Hence, our findings demonstrate how strongly a user is embedded in the Twitter verified user network is highly predictive of their reach in the generic Twittersphere. The scatter plots, acquired regression splines, and confidence intervals are shown in Figure~\ref{fig:fig5}.

\section{Activity Analysis}

Finally, we attempt to characterize the collective tweet activity time series of the network of English verified users. A calendar heatmap for the verified user tweet activity levels over our a collection period of one year can be seen in Figure~\ref{fig:fig6}.  We check for existing auto correlations in the time series using implementations~\cite{b33} of the Ljung-Box and the Box-Pierce portmanteau tests. These tests check for a deviation from the null hypothesis of no auto correlation using a combination of lags, rather than auto correlation with respect to a specific lag. If the $p$ values returned by the test are greater than $0.05$, then the time-lagged correlation cannot be ruled out with a 95\% significance level. We tested for a lag of up to 185 days so as to be able to account for any seasonal auto correlation (quarterly or semi-annual). The Ljung-Box and Box-Pierce test results indicate a maximum $p$ value of $3.81\times10\textsuperscript{-38}$ and $7.57\times10\textsuperscript{-38}$ respectively, thus strongly ruling out any lagged correlation. This countered our initial expectations that there would be a significant auto correlation in a week's lag given that activity rates on Sundays are reliably lower than those on weekdays. Evidence for the same can be seen in the calendar heatmap.

We next inquire whether the activity time series is stationary or not. Existing work on smaller social networks~\cite{b32}, such as Gab, reveal that the activity time series drastically change in response to socio-political events occurring outside the network. Hence, we test for stationarity of the time series - whether a single unchanging distribution produces the series - using an implementation~\cite{b33} of the Augmented Dickey-Fuller test with both a constant term and a trend term. Again, we check for a lag of up to 185 days so as to be able to account for any seasonal changes in distribution (quarterly or semi-annual). For upwards of 250 observations (we have 366) the critical value of the test is $-3.42$ when using a constant and a trend term at the 95\% significance level. If the test statistic value is more negative than the critical threshold, we reject the null hypothesis of a unit root and conclude the presence of stationarity. The ``number of tweets'' time series of the English verified users returns a test statistic of $-3.86$ which is significantly more negative than the critical threshold, thus strongly suggesting stationarity. 

We further confirm this finding using a time series change-point detection mechanism called Pruned Exact Linear Time (PELT)~\cite{b34}. We assume that this time series is drawn from a normal distribution, with mean and variance that can change at a discrete number of change-points. We use the PELT algorithm to maximize the log-likelihood for the means and variances of the underlying distribution with a penalty for the number of change-points. Results from several runs of the algorithm are recorded while cooling down the penalty factor and ramping up the number of change-points. Dates that fall in the change-point list in a significant number of runs of the algorithm are considered viable change-point candidates. We are reliably able to obtain only two change-points - one slightly before Christmas (23\textsuperscript{rd} - 25\textsuperscript{th} December 2017) and another one at the beginning of the summer (around the first week of April). This backs our initial assertion that activity patterns of the English verified Twittersphere are mostly resilient to socio-political events external to the network, especially since our collection period consists of the months leading up to significant global events such as the 2018 FIFA World Cup. This aligns with prior work~\cite{b14} that demonstrates other aspects of the Twitter network, such as topology being resilient to exigent circumstances, such as natural disasters. 

\section{Conclusion and Future Work}
We studied 231,246 English speaking verified Twitter user profiles and the 79,213,811 social connections between them. We characterized their network structure and analyzed user activity patterns of the data collected over a span of one year - July 2017 to June 2018. We observe strong evidence for the presence of a power law in the out-degree and Laplacian eigenvalue distributions in the Twitter network of English verified users. This marks a deviation from findings on the entire Twitter network. Other aspects for which our network deviates from the generic Twittersphere are lower average degree of separation, higher reciprocity, and a large number of attracting components. We also demonstrate how the centrality of a user within this sub-graph is indicative of its influence and reach on Twitter. We have also found that the activity levels of English verified users are largely unaffected by current events extraneous to the network. 

The above-mentioned deviations likely constitute a unique fingerprint for verified users which can be leveraged to discern between a verified and a non-verified user. This can further help evaluate the strength of an unverified user's case for getting verified. These network signatures might also be leveraged for realistic synthetic network generation in the future.

\section*{Acknowledgment}

We thank Language Technologies Research Centre (LTRC) and Precog for their support.

\bibliographystyle{IEEEtran}
\bibliography{refer}
\end{document}